\documentclass[pre,twocolumn,showkeys,showpacs,superscriptaddress]{revtex4-1}
\usepackage{amsmath}
\usepackage{amssymb}
\usepackage{bm}
\usepackage{graphicx}
\usepackage{color}
\usepackage{siunitx}
\newcommand{\re}{{\mathrm e}}
\newcommand{\rd}{\mathrm{d}}
\newcommand{\no}{\hat{n}}
\newcommand{\la}{\langle}
\newcommand{\ra}{\rangle}
\newcommand{\bn}{{\bm n}}
\newcommand{\bW}{{\bm W}}
\newcommand{\bnu}{{\bm \nu}}
\newcommand{\bmu}{{\bm \mu}}
\newcommand{\BES}{{\mathcal{S}}}
\newcommand{\n}[1]{n_#1}
\newcommand{\NR}{N_R}

\begin{document}

\title{A unified theory for excited-state, fragmented, and equilibrium-like Bose condensation in pumped photonic many-body systems}

\date{\today}

\author{Daniel Vorberg}
\email{dv@pks.mpg.de}
\affiliation{Max-Planck-Institut f\"ur Physik komplexer Systeme, N\"othnitzer Str.\ 38, 01187 Dresden, Germany}
\author{Roland Ketzmerick}
\affiliation{Max-Planck-Institut f\"ur Physik komplexer Systeme, N\"othnitzer Str.\ 38, 01187 Dresden, Germany}
\affiliation{Technische Universit\"at Dresden, Institut f\"ur Theoretische Physik, 01062 Dresden, Germany}
\author{Andr\'e Eckardt}
\email{eckardt@pks.mpg.de}
\affiliation{Max-Planck-Institut f\"ur Physik komplexer Systeme, N\"othnitzer Str.\ 38, 01187 Dresden, Germany}

\begin{abstract}
We derive a theory for Bose condensation in nonequilibrium steady states of bosonic quantum gases that are coupled both to a thermal heat bath and to a pumped reservoir (or gain medium),
 while suffering from loss.
Such a scenario describes photonic many-body systems such as exciton-polariton gases.
Our analysis is based on a set of kinetic equations for a gas of noninteracting bosons.
By identifying a dimensionless scaling parameter controlling the boson density,
 we derive a sharp criterion for which system states become selected to host a macroscopic occupation.
We show that with increasing pump power, the system generically undergoes a sequence of nonequilibrum phase transitions.
At each transition a state either becomes or ceases to be Bose selected (i.e.\ to host a condensate):
The state which first acquires a condensate when the pumping exceeds a threshold is
the one with the largest ratio of pumping to loss.
This intuitive behavior resembles simple lasing.
In the limit of strong pumping,
 the coupling to the heat bath becomes dominant so that eventually the ground state is selected,
 corresponding to equilibrium(-like) Bose condensation.
For intermediate pumping strengths, several states become selected giving rise to fragmented nonequilibrium Bose condensation.
We compare these predictions to experimental results obtained for excitons polaritons
 in a double-pillar structure [Phys.\ Rev.\ Lett.\ {\bf 108}, 126403 (2012)]
 and find good agreement.
Our theory, moreover, predicts that the reservoir occupation is clamped at a constant value whenever the system hosts an odd number of Bose condensates.
\end{abstract}

%\pacs{05.70.Ln, 05.30.Jp,66.25.+g,67.85.Jk}
%\keywords{Bose-Einstein condensation, ideal Bose gas, quantum Markov process,
%driven-dissipative quantum systems, non-equilibrium steady state,
%Floquet systems}

%05.30.-d 	*Quantum statistical mechanics*
%05.30.Jp 	Boson systems

%05.70.-a 	*Thermodynamics*
%05.70.Ln 	Nonequilibrium and irreversible thermodynamics

%03.75.-b 	*Matter waves*
%03.75.Hh 	Static properties of condensates; thermodynamical, statistical,
%		and structural properties

%66.25.+g 	*Thermal conduction in nonmetallic liquids*

%67.10.-j 	*Quantum fluids: general properties*
%67.10.Ba 	Boson degeneracy (for ultracold, trapped gases, see 67.85.-d)
%67.10.Fj 	Quantum statistical theory

%67.85.-d 	*Ultracold gases, trapped gases*
%67.85.Bc 	Static properties of condensates
%67.85.Jk 	Other Bose-Einstein condensation phenomena

\maketitle

\section{Introduction}

Lasing \cite{Haken1970, DeGiorgio1970, ManWol1995}
 and Bose-Einstein condensation \cite{AndEnsMatEtAl1995, KenMewAnd1995} have in common that the coherence is build up
 by a macroscopic number of bosons occupying a single mode.
Bose condensation is an equilibrium phenomenon that occurs
  when the temperature is below or the particle density is above a critical value.
The chemical potential increases when increasing the particle density or cooling the system.
In three-dimensional or finite ideal Bose gases, the chemical potential approaches the ground state energy
   so that the occupations of the excited states saturate and the condensate forms in the ground state.
In contrast, a laser is a device that operates far away from thermal equilibrium and emits coherent light due to ``light amplification by stimulated emission of radiation''.
When the gain due to the stimulated emission exceeds the loss in a single-particle state,
 this state acquires a macroscopic occupation.
%In contrast Bose condensation of massive particles like atoms, lasers accumulate (massless) photons and operate far away from equilibrium.

This distinctions between lasing and Bose condensation is a priori not as clear in systems which underly both a coupling to
 a heat bath and a coupling to a pumped particle reservoir.
Exciton-polaritons, hybrid particles of photons and excitons, are a prime example \cite{But2007}.
This has led to the discussion  of how to distinguish between polariton lasing and Bose-Einstein condensation of polaritons \cite{KasSolAndEtAl08, Dev2012,But2007,But2012,ChiGamCar2015,SunWenYooEtAl2017}.
Both phenomena are taken together in the term  ``polariton condensation'' \cite{ByrKimYam2014}.
Only for Bose-Einstein condensation of polaritons, the polariton gas is almost thermalized and approximately obeys the Bose-Einstein distribution.
Furthermore, the nonequilibrium condition allows for excited-state condensation or the coexistence of several condensates \cite{wertz2010spontaneous} and manifests itself in the sensitivity of the condensate density on the pump spot \cite{WouCarCiu2008}.
Also in systems of photons in dye-filled cavities \cite{KlaVewWei2010,SchDamDun2015, Walker2017}, the thermalization competes with gain and loss, leading to a rich phase diagram \cite{KirKee2013, KirKee2015, Hesten2018}. %\todo{add weitz, walker \& nyman}.
Other scenarios, where complex nonequilibrum condensation is expected, are given by periodically-driven systems in contact with a heat bath and by systems exchanging energy with baths of different temperature \cite{Vorberg2013, Vorberg2015, Schnell2017, Schnell2018}.

In this paper we develop a theory for bosonic systems that are coupled to both
 a heat bath and a pumped reservoir while suffering from particle loss (see Fig.~\ref{fig:schematic}).
By identifying a dimensionless scaling parameter controlling the particle density, we derive a sharp criterion for which of the modes are selected to become macroscopically occupied in the high-density limit.
This concept of Bose selection includes lasing-like nonequilibrium condensation in the mode with the largest effective gain,
 equilibrium(-like) ground-state Bose condensation,
 as well as situations in between, where other modes  become selected.
With respect to a variation of the the pump power,
 our theory predicts the following \emph{generic} sequence of nonequilibrum phase transitions:
First, above a threshold, the mode with the largest effective gain becomes macroscopically occupied (corresponding to simple lasing).
Ramping up the pump further, further transitions can occur where single modes acquire or loose macroscopic occupation.
Eventually, a macroscopic occupation of the ground state alone emerges in the limit of strong pumping (resembling equilibrium Bose condensation).
Thus, unless the mode with the largest effective gain coincides with the ground state,
 lasing and equilibrium-like Bose condensation are clearly distinguished by a sequence of transitions.
Our theory, moreover, predicts that the reservoir occupation is clamped at a constant value whenever the system hosts an odd number of Bose selected states.
The special case of a system consisting of only two relevant modes has been investigated in Ref.~\cite{LeyEtAl2016}.

Comparing our theory to experimental data obtained for a system of exciton-polaritons in a double-pillar structure \cite{Galbiati2012},
 we find good agreement.
Furthermore, we point out that the transition to ground-state condensation occurs via an intermediate phase
 where both the mode of largest effective gain and the ground state are Bose selected.

The remaining part of this paper is organized as follows.
In Sec.~\ref{sec:theo_framework}, we introduce the model system which is also sketched in Fig.~\ref{fig:schematic}.
In Sec. \ref{sec:system}, we discuss the specific example of the polariton gas
in a double pillar studied in Ref.~\cite{Galbiati2012}, which in the following will serve as a concrete example illustrating our general results.
The theory for Bose selection in open systems is developed in Sec.~\ref{sec:sel_mech} for a fixed number of reservoir excitations.
The dynamics of the reservoir occupation is then included in Sec.~\ref{sec:res_occup}.
This allows us to derive the generic structure of the phase diagram with respect to pump power.
The selection in the limit of strong pumping where we generically find equilibrium-like Bose condensation is discussed separately in Sec.~\ref{sec:final_selection}.
Subsequently a detailed comparison with the experiment is given in Sec.~\ref{sec:comparison}.
We conclude the paper in Sec.~\ref{sec:conclusion}.

\section{The model}
\label{sec:theo_framework}

\begin{figure}[t]
\centering
\includegraphics[width=\columnwidth]{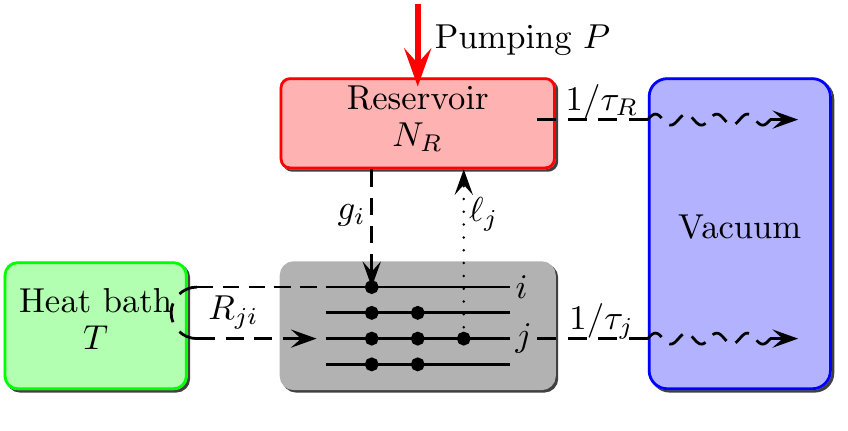}
\caption{
  Sketch of the system (gray) in contact with its environment.
  By exchanging energy with a heat bath of temperature $T$ (green),
   the particles undergo transitions from a state $i$ to another state $j$ at the single-particle rate $R_{ji}$.
  The system looses particles from any state $j$ when the particle escapes the cavity with rate $1/\tau_j$ (blue).
  Furthermore, the system exchanges particles with a pumped reservoir (red),
   which has the same temperature $T$ and hosts $N_R$ particles.
  The corresponding rates for gain and loss processes are $g_i$ and $\ell_j$, respectively.
  The pump and loss rates of the reservoir are $P$ and $\tau_R$, respectively.}
\label{fig:schematic}
\end{figure}

In this paper, we consider the configuration sketched in Fig.~\ref{fig:schematic}. The
system possesses single-particle states $i\in\{0,1,2,\ldots, M-1\}$ ordered by their energy $E_i$.
We describe the dynamics of the mean occupation numbers $n_i=\la\no_i\ra$ by
 the kinetic equations of motion
\begin{align}
  \dot {n}_i=D_{i}^{\mathrm{thermal}}(\{n_j\}) + D_{i}^{\mathrm{gain/loss}}(n_i),
  \label{eq:rate_eq}
\end{align}
which consists of two contributions.

The first contribution compromises all terms describing the intermode kinetics caused by a thermal bath,
\begin{align}
 D_{i}^{\mathrm{thermal}}(\{n_j\})= \sum_{j=0}^{M-1} \big[R_{ij}n_j (n_i+1)-R_{ji} n_i (n_j+1)\big].
 \label{eq:D_inter}
\end{align}
Here, $R_{ji}$ is the single-particle rate for the transition from state $i$ to state $j$ and the dependence on the occupation of the final state, $(n_j+1)$, is due to bosonic enhancement (and captures also stimulated processes).
For a thermal environment of temperature $T$, the rates obey
\begin{align}
 \frac{R_{ji}}{R_{ji}}=\re^{-\beta(E_j-E_i)},
 \label{eq:db}
\end{align}
with inverse temperature $\beta=1/(k_B T)$.
In these processes, the environment exchanges energy with the system only.

The second contribution in the kinetic equations (\ref{eq:rate_eq}) describes all processes where particles are exchanged with the environment,
\begin{align}
 D_{i}^{\mathrm{gain/loss}}(n_i)=G_i (n_i+1) - L_i n_i
  \label{eq:D_open}
\end{align}
Here $G_i(n_i+1)$ is the rate for a transition from the reservoir into the system, where
\begin{align}
 G_i = g_iN_R
\end{align}
is proportional to the reservoir occupation $N_R$ and $g_i$ is the single-particle rate for transition of a reservoir particle to the state $i$.
The term $(n_i+1)$ again reflects the bosonic enhancement.
The loss rate
\begin{align}
 L_i =\ell_i +\tau_i^{-1},
\end{align}
comprises the finite lifetime $\tau_i$ and the rate $\ell_i$ for a transition from the state $i$ to the reservoir.
We assume that the reservoir is given by many states with individual mean occupations much
 smaller than unity, so that bosonic enhancement is negligible within the reservoir.

The reservoir occupation obeys the equation of motion
\begin{align}
\dot{N}_R=&P-\frac{N_R}{\tau_R}+\sum_{i=0}^{M-1}\big[\ell_in_i-N_R g_i(n_i+1)\big],
\label{eq:rate_eq_reservoir}
\end{align}
which, apart from terms describing the particle exchange with the system, is determined by the
 finite lifetime $\tau_R$ of reservoir occupations due to decay processes into states other than the system states and the pumping strength $P$.
In the following, we will investigate the properties of the steady state obtained
 by solving the equations (\ref{eq:rate_eq}) and (\ref{eq:rate_eq_reservoir}) for
\begin{align}
 0=\dot n_i\quad \mathrm{and}\quad 0=\dot N_R.
 \label{eq:steady_state}
\end{align}

In order to derive these kinetic equations of motion, several approximations were made.
First, we neglect the interactions among the particles.
Second, we assume that the bath degrees of freedom are fast compared to the system
 degrees of freedom and that the system-bath coupling is weak compared to the level splitting in the system.
This allows for a description of the coupling to the environment within a Born-Markov approximation and assures that
 that the steady state is described by a density matrix $\rho$ that is diagonal in the
 basis of the system's eigenstates~$i$, $\rho=\sum_\bn p_\bn |\bn\ra\la\bn|$.
Here $|\bn\ra$ denotes the Fock state with occupation numbers $\bn=(n_0,n_1,\ldots, n_{M-1})$.
These approximations lead to a master equation for the probabilities $p_\bn$ for finding the system
 in the Fock states $|\bn\ra$.
%Note, that these approximation may fail in continuous systems since a weak coupling limit requires finite level spacing.
Finally, the nonlinear kinetic equations of motion (\ref{eq:rate_eq}) for the mean occupations $n_i=\la\no_i\ra$
 are obtained by approximating $\la\no_i\no_j\ra\approx \la\no_i\ra\la\no_j\ra$ (see, e.g., Ref.~\cite{Vorberg2015}).

Note that the assumption that the level splitting is large compared to the system bath coupling is challenged in the limit of large systems approaching a continuous spectrum.
Relaxing this condition, one can obtain equations of motion also for the off-diagonal elements of density matrix, which can be solved numerically.
However, this is not the aim of the present paper.
Instead, we keep this assumption, which is well justified, e.g., for the experimental system studied in Ref.~\cite{Galbiati2012} (which we will use to illustrate our findings).
This will allow us to make analytical statements about the nonequilibrium phase diagram of the system.

\section{Example system: photonic molecule}
\label{sec:system}

\begin{figure}[t]
\centering
\includegraphics[width=\columnwidth]{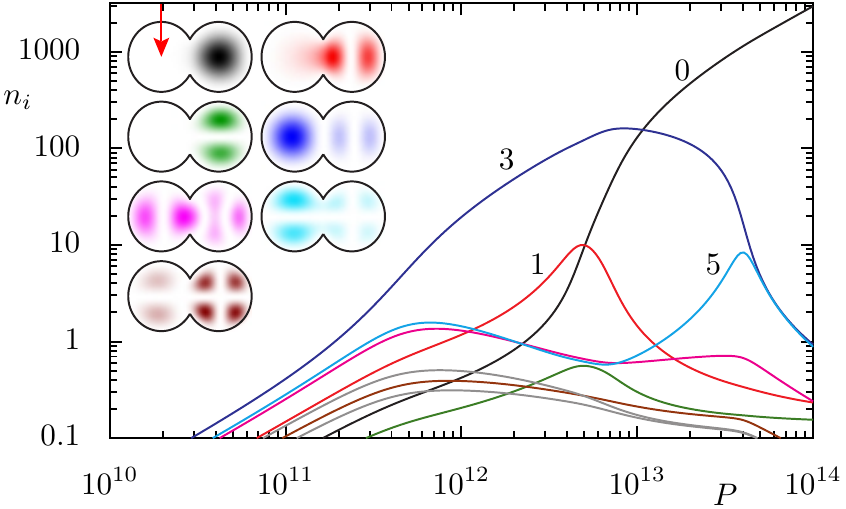}
\caption{
  Input-output characteristics for a double-pillar system.
  The shown dependence of the mode occupation numbers $n_i$ on the pump power $P$ (in arbitrary units) was obtained by Eqs.~(\ref{eq:rate_eq}), (\ref{eq:rate_eq_reservoir}), and (\ref{eq:steady_state}).
  The inset shows the lowest eigenmodes of the system.
  Viewed from the top, the double-pillar system looks like two overlapping disks.
  The pumping in the left pillar is indicated by the red arrow.
  The probability densities $|\psi_i(x, y)|^2$ are shown in the respective colors as in the main figure. They are ordered energetically row by row from top to bottom, with
  $i=0$ (black), 1 (red), 2 (green), 3 (blue), 4 (pink), 5 (cyan), 6 (brown).
  The ground state (black, top left) is localized in the right column as a result
   of the repulsive potential created by the reservoir in the left column.}
\label{fig:polariton_exp}
\end{figure}

In order to illustrate and test the general theory that will be developed in the following sections, we will apply it to a concrete system given by exciton-polaritons in a two-dimensional double-pillar geometry [see inset of Fig.~\ref{fig:polariton_exp}].
Such a ``photonic molecule'' has recently been investigated experimentally and numerically in Ref.~\cite{Galbiati2012}.
The system is pumped in an asymmetric fashion in the left pillar.
Therefore, the reservoir particles (occupying highly excited states overlapping with the low-energy
 states defining the system \cite{DengEtAl10,CarusottoCiuti13}) are located in the left column.
Due to repulsive interactions, the reservoir excitions give rise to a mean-field potential energy shift in the left pillar.

We model the system by the potential $V(x,y)= V_\text{dp}(x,y)+V_\text{off}(x,y)$.
The potential of the double pillar $V_\text{dp}(x,y)$ vanishes for $(x\pm a)^2+y^2<(d/2)^2$ and is infinite elsewhere,
 where the diameter and the distance of the columns is given by $d=4 \mu \mathrm{m}$ and $a=3.46\mu\mathrm{m}$, respectively [see overlapping disks in insets of Fig.~\ref{fig:polariton_exp}].
The mean-field shift is mimicked by an offset $V_\text{off}(x,y)$, which takes the values $0$ and
 $2.05 \mathrm{meV}$ for $x>0$ and $x\le 0$, respectively.
The lowest eigenstates are shown in the inset of Fig.~\ref{fig:polariton_exp} ordered by their energy (the color code of
 this figure will be used throughout this paper).
The derivation of the rates follows Ref.~\cite{Galbiati2012} and is discussed in appendix \ref{app:polartion_rates}.

The input-output characteristics of the modes are shown in Fig.~\ref{fig:polariton_exp}.
The steady-state occupations versus the pump power $P$ are computed
 from the kinetic equations (\ref{eq:rate_eq}) and (\ref{eq:rate_eq_reservoir}).
We have taken into account the ten states with lowest energy.
The results describe the experimental findings of Ref.~\cite{Galbiati2012}:
For small pumping all states are occupied weakly.
Above a threshold, the third excited state (blue) acquires a large occupation corresponding to Bose selection.
We will see that this state is singled out by the largest effective gain $G_i/L_i$, i.e., by its overlap with the reservoir.
%This state is singled out by its lowest energy from all states which have a significant spatial overlap with the reservoir.
In the limit of strong pumping, eventually a condensate is formed in the ground state (black).
One can also observe that between these regimes, more than one state is selected.
In the following sections, we will derive an analytical theory that shows that the behavior visible in Fig.~\ref{fig:polariton_exp} reflects generic properties of pumped bosonic systems described by Eqs.~(\ref{eq:rate_eq}) and (\ref{eq:rate_eq_reservoir}).

\section{Bose selection}
\label{sec:sel_mech}
\begin{figure}[t]
\centering
\includegraphics[width=\columnwidth]{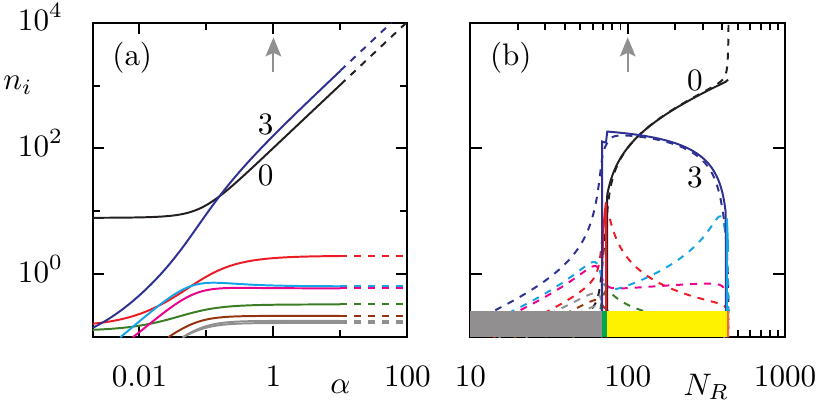}
\caption{
 (a) Steady state of Eq.~(\ref{eq:rate_eq}) for fixed reservoir occupation $N_R=100$, versus the parameter $\alpha$.
 In the limit of large $\alpha$, the ground state (black) and the third excited state (blue) acquire large occupations $\propto \alpha$
  while the occupations of all other modes saturate.
 (b) Mean occupations $n_i$ versus the reservoir occupation $N_R$.
 The exact solution of the rate equation (dashed lines) is compared with the asymptotic theory for Bose-selected states (solid lines).
 Increasing the reservoir occupation $N_R$ triggers transitions between different sets of Bose-selected states.
 Each set is indicated by a unique color at the bottom (gray for $\BES=\{\}$, green for $\BES=\{1,3\}$, yellow for $\BES=\{0,3\}$, and orange for $\BES=\{0,5\}$).
 Above a maximum value of $N_R$, no steady-state solution is found.}
\label{fig:n_vs_g_and_alpha}
\end{figure}

In order to analyze the limit of strong driving, we introduce the scaling factor $\alpha$ multiplying the rates for gain and loss, i.e., the term $D_{i}^{\mathrm{gain/loss}}$ [cf.~Eq.~(\ref{eq:rate_eq})]
\begin{align}
  \label{eq:scaling}
  \dot {n}_i=D_{i}^{\mathrm{thermal}}(\{n_j\}) + \alpha D_{i}^{\mathrm{gain/loss}}(n_i),
\end{align}
Later, we will set $\alpha=1$.
Figure~\ref{fig:n_vs_g_and_alpha}(a) shows the mean occupations $n_i$ versus the parameter $\alpha$
 computed from Eq.~(\ref{eq:scaling}) with the gain parameters $G_i$ obtained for the fixed reservoir occupation $N_R=100$.
In the asymptotic regime of large $\alpha$, one can observe that the states clearly separate into two groups:
The first group is formed by states whose occupations increase linearly with $\alpha$ and we refer to these states in the following as Bose-selected states
 \footnote{As we will explain in detail below, this effect is a generalization of the phenomenon of Bose selection recently discussed in nonequilibrium steady states of driven-dissipative Bose gases not exchanging particles with their environment \cite{Vorberg2013}.}.
The occupations of all other states, the nonselected states, saturate in the limit of large $\alpha$.

Among both groups the relative occupations do not change anymore.
When the system has already reached this asymptotic regime for $\alpha=1$,
 we can identify the selected states as those single-particle states hosting a Bose condensate.
Thus, the scaling parameter $\alpha$ allows us to give a sharp distinction between condensed and noncondensed modes.
Moreover, as we will see below, essential properties of the system can be described analytically within a simple asymptotic theory,
 obtained from the leading-order terms of a systematic expansion in powers of $1/\alpha$.
The dashed lines in Fig.~\ref{fig:n_vs_g_and_alpha}(a) are computed within such an asymptotic theory.
Here, this theory provides a good approximation already for $\alpha=1$.

In order to derive the asymptotic theory, we expand the mean occupations in a power series with
respect to $\alpha^{-1}$,
\begin{align}
 n_i=\nu_i\alpha+\nu_i^{(1)}+\nu_i^{(2)}/\alpha+\ldots.
 \label{eq:series_expansion}
\end{align}
After plugging this ansatz into the rate equatons \eqref{eq:rate_eq}, we find
\begin{align}
 0=\alpha^2&\left[\sum_j(R_{ij}-R_{ji})\nu_i\nu_j+(G_i- L_i)\nu_i \right]\nonumber\\
  +\alpha&\left[\sum_j(R_{ij}\nu_j-R_{ji}\nu_i+(R_{ij}-R_{ji})(\nu_j\nu_i^{(1)}+\nu_j^{(1)}\nu_i))\right.\nonumber\\
 &\left.+( G_i- L_i)\nu_i^{(1)} + G_i \vphantom{\sum_j}\right]
 +\ldots.
 \label{eq:eom_series_eqxansion}
\end{align}
From the vanishing leading-order coefficient, we obtain
\begin{equation}
 0=\nu_i\left(\sum_{j=1}^M A_{ij}\nu_j+ W_i\right)\!,
 \label{eq:naive_app}
\end{equation}
where we have introduced the short-hand notations
\begin{equation}
 A_{ij}=R_{ij}-R_{ji}
\end{equation}
for the asymmetries of the intermode rates and
\begin{equation}
 W_i= G_i- L_i
\end{equation}
for the gain-loss asymmetries.

The set of equations (\ref{eq:naive_app}) possesses solutions of the form
\begin{equation}
 \left\{\begin{array}{ll}
  \sum_{j\in\BES} A_{ij}\nu_j+W_i=0 &\text{ for }  i\in\BES\\
  \nu_i=0 &\text{ for }  i\notin\BES \,.
 \end{array}
 \right.
 \label{eq:naive_guess}
\end{equation}
Here we have introduced the (yet to be determined) set of selected states $\BES$.
The form (\ref{eq:naive_guess}) explains already the observation of Fig.~\ref{fig:n_vs_g_and_alpha}(a)
 that the states of the system separate into two groups,
 selected states $i\in\BES$, whose occupation increase linearly with $\alpha$
 and nonselected states whose occupations saturate for large $\alpha$.

The leading contribution to the occupations of the nonselected states is given by the
 coefficients $\nu_i^{(1)}$.
They can be obtained by requiring the terms proportional to $\alpha$ of Eqs.~(\ref{eq:eom_series_eqxansion}) to vanish and read
\begin{align}
 \nu_i^{(1)}=-\frac{1}{\mu_i}\left[\sum_{j\in\BES}R_{ij}\nu_j+ G_i\right]
 \quad \forall i\notin\BES ,
 \label{eq:naive_guess_non_sel}
\end{align}
with
\begin{align}
  \mu_i = \sum_{j\in\BES} A_{ij}\nu_j+W_i.
\end{align}

We have still not answered which of the states become selected.
The set of selected states $\BES$ is determined by the physical condition of nonnegative (asymptotic) mean occupations,
 i.e., $n_i\simeq\nu_i>0\ \forall i\in\BES$ and $n_i\simeq\alpha\nu^{(1)}_i>0\ \forall i\notin\BES$.
For the nonselected states, the sign of the occupations is determined by the denominator of
 Eq.~(\ref{eq:naive_guess_non_sel})
 since $R_{ij}$, $\nu_j$, and $G_i$ appearing in the numerator are positive.
Thus the second condition is equivalent to $\mu_i<0\ \forall i\notin\BES$.
In conclusion, the set $\BES$ must fulfill the selection condition
\begin{equation}
 A\bnu +\bW =\bmu
\mbox{ with }
 \left\{\begin{array}{ll}
 \nu_i\ge 0  \land \mu_i=0 & \text{ for }\,i\in\BES \\
 \nu_i=0 \land \mu_i\le 0 &\text{ for }\,i\notin\BES,
 \end{array}\right.
 \label{eq:selection_criterion}
\end{equation}
where $\bnu$ and $\bW$ are vectors with elements $\nu_i$ and $W_i$ and $A$ is the rate-asymmetry matrix with elements $A_{ij}$.
As shown in detail in appendix~\ref{sec:implications}, from the selection criterion (\ref{eq:selection_criterion}), we can immediately draw two conclusions:
 (i) Without fine-tuning of the gain-loss asymmetries $\bW$, the set of selected states contains an even number of states;
 (ii) the set of selected states is unique.
In the next section we will show that implication (i) does not exclude situations with an odd
 number of Bose condensates over extended intervals of the pump power.

\begin{figure*}[t]
\centering
\includegraphics{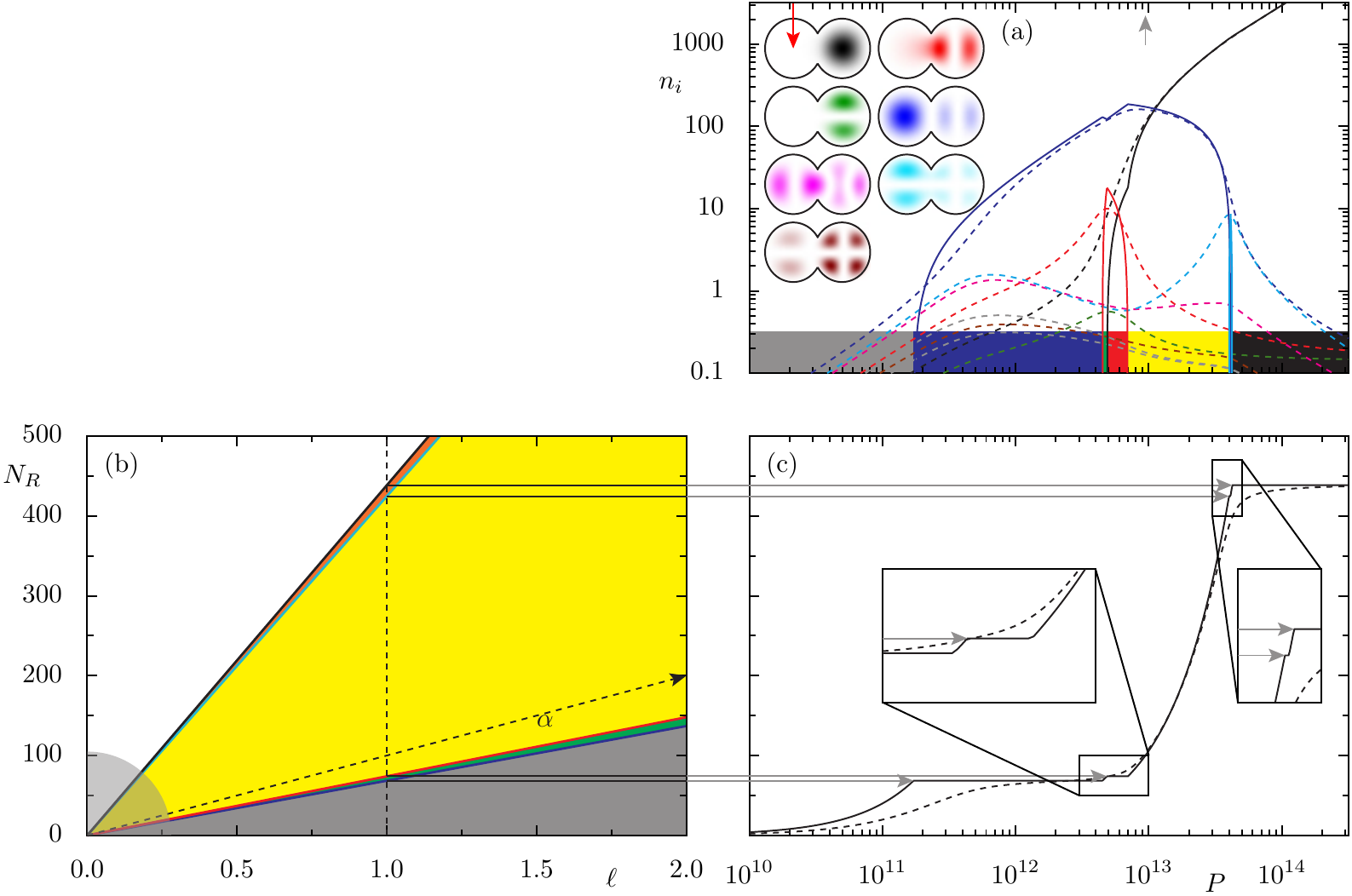}
\caption{
 (a) Mean occupations $n_i$ versus pumping strength $P$ (dashed lines: exact solution of the rate
  equations, solid lines: asymptotic theory for selected states).
 The colors at the bottom refer to the color code of the phase diagram.
 (b) Phase diagram in the plane spanned by the reservoir occupations $N_R$ controlling the gain $G_i=N_Rg_i$
  and the scaling parameter $\ell$ controlling the loss $\ell L_i$.
  The phase diagram is computed within the asymptotic theory, which is not valid for small gain and loss, as indicated by the gray-shaded quarter circle at the origin.
 The colored areas correspond to different sets of selected states which are (going along the vertical dashed line from bottom to top)
  $\BES=\{\}$ (gray area), $\BES=\{3\}$ (blue line),  $\BES=\{1, 3\}$ (thin green area), $\BES=\{0,1,3\}$ (red line), $\BES=\{0,3\}$ (yellow area), $\BES=\{0,3,5\}$ (turquoise line), $\{0,5\}$ (orange area), and $\{0\}$ (black line).
 In the white region, no steady-state solution is found.
 Figures~\ref{fig:n_vs_g_and_alpha}(a) and (b) show cuts through this phase diagram
  along the dashed vertical and radial line, respectively.
 Indeed, increasing $\ell$ and $\NR$ proportionally to each other is equivalent to increasing $\alpha$.
 (c) Reservoir occupation $N_R$ versus pumping strength $P$ for $\ell=1$
  (dashed lines: exact solution, solid: line asymptotic theory).
 At values of $N_R$ corresponding to a transition in the phase diagram (b), plateaus are formed, where the reservoir occupation $N_R$ (c) is clamped, i.e., does not increase with
respect to $P$. At these plateaus, the set of selected states contains an odd number of states,
which we label by the color of the transition lines in (b).
Two of these plateaus are very small as shown in the two zoom boxes.
The curve (c) relates the phase diagram (b)
to the plot (a). It is a generic feature that the first state to become selected when the
pumping is increased, here state $3$, is singled out by the largest gain-to-loss ratio and that for
strong pumping the ground state $0$ is selected.}
\label{fig:vs_P}
\end{figure*}

A variation of the reservoir occupation $N_R$ and thus of the asymmetry $W_i=g_iN_R-\ell_i-1/\tau_i$
 can trigger nonequilibrum phase transitions from one set of selected states to another one.
This is shown in Fig.~\ref{fig:n_vs_g_and_alpha}(b).
At each transition, two states change their classification of being either selected or nonselected
 so that an even number of states is selected after the transition [as required by criterion (\ref{eq:selection_criterion})].
Within the asymptotic theory (solid lines), the transitions are discontinuous (first-order-like),
 as the mean occupations undergo finite jumps.
Below a threshold value of the reservoir occupation $N_R$,
 no state is selected,
 while above a maximum value of $N_R$, no steady state solution is found.

The phase diagram shown in Fig.~\ref{fig:vs_P}(b) shows how the set of selected states depends on the reservoir occupation $\NR$ and
the parameter $\ell$ describing a global scaling of the loss rates,
\begin{align}
 L_i \to \ell L_i.
\end{align}
We have already seen two cuts through this phase diagram.
One the one hand, Fig.~\ref{fig:n_vs_g_and_alpha}(a) corresponds to a radial cut.
In this direction no transitions occur but the separation between the occupations of selected and nonselected states increases.
On the other hand, Figure~\ref{fig:n_vs_g_and_alpha}(b) corresponds to a vertical cut through the phase diagram shown in Fig.~\ref{fig:vs_P}(b) along the dashed line.
We encounter the following sequence with increasing $N_R$: After initially no state is selected, a phase with two selected states (corresponding to a fragmented condensate in the ground state and the third excited state) follows, before eventually for large $N_R$ a regime without steady state solution is found.
Furthermore, two phases appear in between (green and orange) which span only over small intervals of the reservoir occupation $N_R$.

\section{Impact of reservoir dynamics}
\label{sec:res_occup}
Now we release the restriction of a fixed reservoir occupation $N_R$ by including its dynamics given by Eq.~\eqref{eq:rate_eq_reservoir}.
Figure \ref{fig:vs_P}(a) and (c) show the mode occupations $n_i$ and reservoir occupation $\NR$ versus the pump power $P$, respectively.
The occupations of selected states given by the asymptotic theory (solid lines) provide a good approximation as compared to the exact solution of kinetic Eqs. (\ref{eq:rate_eq}) and (\ref{eq:rate_eq_reservoir}) [dashed lines].
We can observe that varying the pumping triggers transitions where one state either starts or ceases to be selected.
Thus the set of selected states changes with increasing pumping.

With the pumping $P$ also the reservoir occupation $\NR$ increases.
However, whenever a phase boundary is crossed in Fig.~\ref{fig:vs_P}(b), the reservoir occupation is clamped at a certain value over a finite interval of pump powers $P$ [plateaus shown in \ref{fig:vs_P}(c)].
These phase boundaries, which are represented by colored radial lines in the phase diagram of Fig.~\ref{fig:vs_P}(b), correspond to situations where an odd number of states are selected.
The critical values of the reservoir occupation $N_R$ at which it is clamped correspond to a fine-tuned values for the gain-loss asymmetry $\bW$.
Here odd number of states is selected over the finite interval of pumping strengths corresponding to the plateau.
This behavior is related to the fact that the transitions in the phase diagram are discontinuous with respect to the reservoir occupation $N_R$.
The total number of particles in the system $N=\sum_in_i$ undergoes a jump $\Delta N=N_>-N_<$
 when $N_R$ is varied across a transition from a value $N_<$ directly before the transition
 to a value $N_>$ directly after the transition.
Thus, whenever the reservoir occupation $N_R$ reaches a transition, further increase of $P$ will not alter $N_R$,
 while $N$ continuously changes from $N_<$ to $N_>$.
In this regime, which bears resemblance to a coexistence phase (e.g.\ between liquid and gas), an odd number of particles is selected.

The theory that we will develop below reveals that several of the features that we see in Fig.~\ref{fig:vs_P} correspond to the following generic behavior:
\begin{itemize}
  \item With respect to the pump power, the system undergoes a sequence of transitions,
    where at each transition either a nonselected state becomes selected or
    a selected state becomes nonselected.
  \item Whenever an odd number of states (e.g.\ a single state) is selected,
   the reservoir occupation is clamped with respect to the pump power.
  \item The first state to become selected when the pump power is switched on is the one with largest ratio of gain to
   loss, $g_i/L_i$ (third state in the present example). This behavior corresponds to simple lasing.
  \item In the limit of strong pump power $P$, the selected states are determined by the rates $R_{ij}$ alone.
   When the rates obey condition (\ref{eq:db}), this implies a single condensate in the ground state.
\end{itemize}
The above conditions imply also:
\begin{itemize}
  \item There is no direct transition from lasing in an excited state to ground-state condensation with respect to the pump power.
  There must be at least one intermediate phase where two modes are selected.
\end{itemize}

In order to derive these statements, let us now include the reservoir dynamics into the asymptotic theory.
In leading order $\alpha^{-1}$ and for the steady state $\dot N_R=0$,
 the rate equation (\ref{eq:rate_eq_reservoir}) reduces to
\begin{align}
 0=P-\frac{N_R}{\tau_R}+\sum_{i\in\BES}(\ell_i-N_R g_i)\nu_i,
 \label{eq:pump_asy}
\end{align}
where only the selected states with nonzero $\nu_i$ contribute.

\subsection{First transition}

In the limit of small pumping, when no state is selected yet, $\BES=\{ \}$,
 the reservoir occupation increases linearly with the pumping, $N_R=P\tau_R$ [Eq.~\eqref{eq:pump_asy}].
In this case, the criterion~\eqref{eq:selection_criterion} is fulfilled by
\begin{equation}
 \nu_i=0
\end{equation}
and
\begin{equation}
 \mu_i= W_i = N_R g_i - L_i <0.
\end{equation}
Thus, when the pumping $P$ is increased, the first transition occurs when $W_k$ becomes zero for a state $k$.
This occurs at
\begin{equation}
 N_R^*=L_k/g_k
\end{equation}
respectively
\begin{equation}
P^*=L_k/(g_k\tau_R)
\end{equation}
and the state $k$ is singled out by the largest gain-to-loss ratio, $g_k/L_k > g_i/L_i\ \forall i\neq k$.
At this transition, the state $k$ becomes selected.
In our example system, we have $k=3$ [dark blue line in Fig.~\ref{fig:polariton_exp}, \ref{fig:n_vs_g_and_alpha}, and \ref{fig:vs_P}(a)].

\subsection{Second transition}

After the transition, we have $\BES=\{k\}$.
The selection criterion (\ref{eq:selection_criterion}) requires $\mu_k=0$,
 implying that the reservoir occupation $N_R$ is clamped at the critical value $N^*_R$.
This is a fine-tuned situation which corresponds to the blue line in the phase diagram shown in Fig.~\ref{fig:vs_P}(b).
The curve $N_R(P)$ forms a plateau when $P$ increases above $P^*$.
The occupations of the selected state are then determined by Eq.~(\ref{eq:pump_asy}), giving
\begin{align}
 \nu_k = \frac{P-N^*_R\tau_R^{-1}}{L_k N_R^*-\ell_k}
\end{align}
so that
\begin{align}
 \mu_i = A_{ik}\nu_k + N_R^*  g_i - L_i\ \forall i\neq k.
\end{align}

The selected state occupation $\nu_k$ increases with respect to $P$ and the next transition occurs
 when $\mu_q$ becomes zero for another state $q\ne k$ at a critical pumping strength $P^{**}$.
Note that for thermal rates $R_{ij}$, which obey relation (\ref{eq:db}),
 one has $A_{ij}=R_{ij}-R_{ji} > 0$ for states $i$ with energy $E_i<E_j$.
In this case, the next state $q$ to become selected will have a lower energy $E_q<E_k$.
This implies that no further transition will occur if and only if the selected state $k$ is the ground state.
In our example system, we have $q=1$ [red line in all Fig.~\ref{fig:polariton_exp}, \ref{fig:n_vs_g_and_alpha}, and \ref{fig:vs_P}(a)].

\subsection{Third transition}

After the second transition, the set of selected states is given by $\BES=\{k,q\}$ and the
selection criterion \eqref{eq:selection_criterion} is solved by
\begin{align}
A_{kq}\nu_q + W_k=0 \quad\Rightarrow \quad &
		\nu_q = -\frac{W_k}{A_{kq}} =\frac{g_k N_R-L_k}{A_{kq}},
\nonumber\\
A_{qk}\nu_k + W_q=0 \quad\Rightarrow \quad &
		\nu_k = -\frac{W_q}{A_{qk}} =\frac{-g_q N_R-L_k}{A_{qk}}.
\end{align}
This solution does not require fine tuning of the reservoir occupation $N_R$.
Instead, the reservoir occupation is determined by Eq.~(\ref{eq:pump_asy}) and increases with $P$,
\begin{align}
 N_R=\tau_R P+\tau_R \sum_{i=k,q}(\ell_i-N_R g_i)\nu_i.
\end{align}
We can infer that $\nu_q$ increases and $\nu_k$ decreases with increasing $P$.
Moreover, for a general state $i$, we have
\begin{align}
 \mu_i = A_{ik}\nu_k+A_{iq}\nu_q + W_i .
\end{align}
When the pumping $P$ is increased, the next transition occurs when either $\mu_{p}=0$
 for a state $p\notin \{q,k\}$ or $\nu_k=0$ (depending on what happens first).
In this transition either the state $k$ ceases to be selected, so that after the transition $\BES=\{q\}$
 or the state $p$ becomes selected, so that after the transition $\BES=\{k,q,p\}$, respectively.
In our example system, the second case occurs with the third selected state being given by the ground state, $p=0$
 [black line in Figs.~\ref{fig:polariton_exp}, \ref{fig:n_vs_g_and_alpha}, and \ref{fig:vs_P}(a)].

\subsection{Transitions from even to odd numbers of selected states}

After having gained some intuition from discussing the first transitions occurring when the pumping $P$ is switched on,
 let us now discuss the general case.
For that purpose, we assume that the system has entered a phase characterized by the set of selected states $\BES$ when the pumping $P$ was
 raised above the critical value.
At this transition, a state either starts or ceases to be selected.
For this situation, we wish to compute the occupation numbers and to determine the critical pumping strength $P^{*}$
 at which the next transition occurs, because a state $k$ either
 leaves or enters the group of selected states.
We will treat phases with an even and an odd number of selected state separately.
The difference between both cases results from different properties of the matrix $A_\BES$,
 defined as the projection of the rate-asymmetry matrix to the subspace of the selected states.
Whereas for an even number of selected states, this matrix is generically (i.e.\ without fine tuning) invertible,
  it is singular for an odd number of selected states since skew-symmetric matrices posses an eigenvalue zero when acting in an odd-dimensional space.

We start with the case where the number of selected states is even.
Here the matrix $A_\BES$ is invertible, $A_\BES A_\BES^{-1}=1$.
Therefore, the solution fulfilling the criterion (\ref{eq:selection_criterion}) is given by
\begin{align}
\nu_i(N_R) &=\underbrace{\sum_{j\in\BES}\left(A^{-1}_{\BES}\right)_{ij}L_i}_{\nu_i^A} -N_R\underbrace{ \sum_{j\in\BES}\left(A^{-1}_{\BES}\right)_{ij}g_j}_{-\nu_i^B}\ \forall i\in\BES
\label{eq:nu_A}\\
\end{align}
and
\begin{align}
\mu_i(N_R) &=\underbrace{\textstyle \sum_{j}A_{ij}\nu_j^A-L_i}_{\mu_i^A}+N_R\underbrace{\textstyle(g_i+\sum_jA_{ij}\nu_j^B)}_{\mu_i^B}.
\label{eq:mu_A}
\end{align}
The reservoir occupation follows from Eqs.~\eqref{eq:rate_eq_reservoir} and increases linearly
 with the pumping,
\begin{align}
 %P={N_R}/{\tau_R}+\sum_i\nu_i(\ell_i-N_R g_i)
 N_R = \frac{P+\sum_i\nu_i\ell_i}{1/\tau_R+\sum_i\nu_ig_i}.
\end{align}
Thus, also the parameters $\nu_i$ and $\mu_i$ depend linearly on the pumping $P$.
When the pumping is increased, a transition is triggered by a state $k$ for which either $\nu_k=0$ when it is selected or $\mu_k=0$ when it is nonselected.
We denote the critical reservoir occupation by $N_R^{*}$ and the corresponding critical pumping strength by $P^{*}$.
From Eqs.~\eqref{eq:nu_A} and (\ref{eq:mu_A}), we find the critical reservoir occupation to be given by
\begin{align}
 N_R^{*}=\min\{N_{R}^i|N_{R}^i>N_R'\},\; N_{R}^i=
  \begin{cases}-{\nu_i^A}/{\nu_i^B}&\mbox{if }i\in\BES\\
   -{\mu_i^A}/{\mu_i^B}&\mbox{if }i\notin\BES.\\
  \end{cases}
\end{align}
Here $N_R'$ denotes the reservoir occupation found at the previous transition (which is zero if it is the first threshold).
The state $k$ changes its classification, i.e., becomes nonselected (selected)  when it was selected (nonselected) before the transition.
There is always a state which eventually triggers a transition since at least one of the $\nu_i^B$ for a selected state $i$ is negative. %{\color{red}{why}}
The number of selected states after the transition is odd.

\subsection{Transitions from odd to even numbers of selected states}

In a phase where the number of selected states is odd, the matrix $A_{\BES}$ is singular.
Thus the equation
\begin{align}
 \sum_{j\in\BES} A_{ij}\nu_j + W_j=0  \quad \forall i\in\BES
\end{align}
appearing in the criterion~(\ref{eq:selection_criterion}) has a family of solutions which is given by
\begin{equation}
 \nu_i=\nu_i^A+\lambda\nu_i^B\quad \forall i\in\BES,
 \label{eq:nu_B}
\end{equation}
and parametrized by $\lambda$.
Here $\nu_i^A$ denote the occupations of selected states $i\in\BES$ at the previous transition (which is of the form discussed above)
and $\nu_i^B$ is the homogeneous solution of
\begin{align}
 \sum_{j\in\BES} A_{ij}\nu_j^B=0\quad \forall i\in\BES.
 \label{eq:homogeneous_solution}
\end{align}
We normalize the vector $\nu_i^B$ such that $\sum_i\nu_i^B=\pm 1$, with the sign chosen such
 that the occupation of the state $k$ which triggered the previous transition is positive for sufficiently small $\lambda>0$.
The parameters $\mu_i$ are given by
\begin{align}
 \mu_i=\mu_i^A+\lambda\mu_i^B
 \label{eq:mu_B}
\end{align}
with
\begin{align}
 \mu_i^B = \sum_{j}A_{ij}\nu_j^B.
 \label{eq:mu_homogeneous_solution}
\end{align}
The parameter $\lambda$ plays a role analogously to the reservoir occupation in regions with an even number of selected states.
When the pumping $P$ is increased (with respect to the critical pumping where this phase was entered)
 the solution is described by positive $\lambda>0$, while the reservoir occupation is clamped at the critical reservoir occupation $N_R^*$
 since the family of solutions requires the fine-tuned parameters $W_i(N_R^*)$.
Ensuing from Eq.~(\ref{eq:pump_asy}), the parameter $\lambda$ depends linearly on $P$,
\begin{align}
  \lambda = \frac{P - \frac{N_R}{\tau_R}
  		- \sum_i\nu_i^A(N_R^* g_i-\ell_i)}{\sum_i\nu_i^B(N_R^* g_i-\ell_i)}.
\end{align}
Thus also the $\nu_i$ and the $\mu_i$ change linearly with $P$.
When the pumping is increased, another transition may occur at a critical pumping strength
 $P^{**}$ when a state $q$ reaches either $\nu_q=0$ if the state is selected or $\mu_q=0$ if the state is nonselected.
This state is given by
\begin{align}
 q=\underset{i}{\operatorname{argmin}} \{\lambda_i|\lambda_i>0\},\quad \lambda_i
	=\begin{cases} -{\nu_i^A}/{\nu_i^B}&\mbox{ if }i\in\BES\\
			-{\mu_i^A}/{\mu_i^B}&\mbox{ if }i\notin\BES.\\
\end{cases}
\label{eq:alpha_star}
\end{align}
If such a state $q$ exists, it changes at $P^{**}$ its classification and the number of selected states is even and the process is repeated.
However, there is no state which triggers a transition when $\lambda_i<0\ \forall i$. In this case no further transition occurs as discussed below.

\section{Bose selection in the limit of strong pumping}
\label{sec:final_selection}

\begin{figure}[t]
  \centering
  \includegraphics[width=\linewidth]{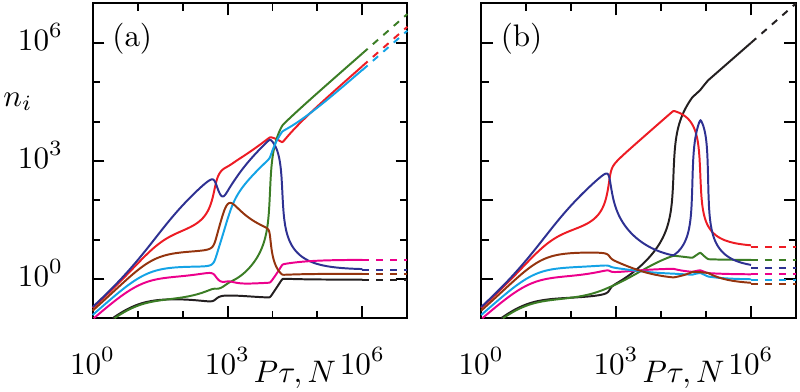}
  \caption{
    Comparison of mean-occupations $\n{i}$ between a system exchanging energy and particles with its environment (solid lines, versus scaled pump power $P\tau$) and the corresponding system without particle exchange (dashed lines, versus total particle number $N$).
    (a) Rates are given by independent uniformly distributed random numbers, $R^{(a)}_{ij}\sim \mathcal{U}(0,0.001)$.
    Here several states are selected in the limit of strong pumping and large particle number, respectively.
    The mean occupation numbers $\n{i}$ of the open and closed system approach each other in the limit of strong pumping $P$ and large total particle numbers $N$, respectively. Thus, the same set of states is Bose selected.
    (b) The thermal rates are obey condition (\ref{eq:db}) and are given by Eq.~(\ref{eq:thermal_rates}).
    In the limit of strong pumping $P$ for the open system and of large total particle numbers $N$ for the closed system, only the ground state (black line) is selected.
    (a,b) The loss rates are state independent, $\tau_i=\tau=1, \ell_i=0\ \forall i$, and the gain rates are independent uniformly distributed random numbers, $G_i\sim \mathcal{U}(0, 1)$.
    The number of particles in the open system approaches $N\approx P \tau$ since almost all particles pumped at rate $P$ are transferred to the system and stay there for the lifetime $\tau$.
      }
  \label{fig:open_vs_close}
\end{figure}

The system reaches a final Bose selection in the limit of strong pumping.
Thus increasing the pumping does not change the set of selected states anymore
when for an odd number of selected states $\lambda_i\le0\ \forall i$ [Eq.~(\ref{eq:alpha_star})].
We denote this final set of selected states by $\BES_f$.
This final set of selected states is determined by the intermode rates $R$ only via the selection condition
\begin{equation}
 A\bnu  =\bmu \text{  with }
 \left\{\begin{array}{ll}
 \nu_i\ge 0  \land \mu_i=0 & \text{ for }\,i\in\BES_f \\
 \nu_i=0 \land \mu_i\le 0 &\text{ for }\,i\notin\BES_f.
 \end{array}\right.
 \label{eq:final_criterion}
\end{equation}
This condition follows from Eqs.~(\ref{eq:homogeneous_solution}) and (\ref{eq:mu_homogeneous_solution})
 where we have dropped the superscript $B$ at the coefficients $\nu_i$ and $\mu_i$.
The constraint that no further transition occurs, i.e.\ $\lambda_i\le 0\ \forall i$ in Eq.~(\ref{eq:alpha_star}),
 implies that $\nu_i^B\ge 0\ \forall i\in \mathcal{S}$ and $\mu_i^B<0\ \forall i\notin\mathcal{S}$ according to criterion (\ref{eq:alpha_star}).
% implies also $\nu_i^A\ge 0$ for $i\in \mathcal{S}$ and $\mu_i^A<0$,
% if $i\notin\mathcal{S}$ for $\lambda_i< 0\ \forall i$.
%Thus, the implies that the right-hand side of Eq. (\ref{eq:nu_B}) is nonnegative and that the left-hand side of Eq. (\ref{eq:mu_B}) is negative,
% so that all occupation numbers are positive.
%This condition follows from Eqs.~(\ref{eq:homogeneous_solution}) and (\ref{eq:mu_homogeneous_solution})
%The latter implies that $\nu_i^B\ge 0$ for $i\in\BES$ and that $\mu_i^B\le 0$ for $i\notin\BES$.
%Note that $\nu_i^A>0\ \forall i\in\BES$ and $\mu_i^A<0\ \forall i\notin\BES$ holds since this requires nonnegative occupations in the modes.
The condition (\ref{eq:final_criterion}) for the set $\BES_f$ differs from the
selection criterion for a general phase, given by Eq.~(\ref{eq:selection_criterion}) by the absence of the gain-loss
difference $\bm W$.
Thus, the final set of selected states depends on the rates $R_{ij}$ only
 and is independent of the gain and loss rates $G_i$ and $L_i$.

This result might appear counterintuitive
 since the final set of selected states $\BES_f$ is reached in the limit of strong pumping.
It can be explained, however, by the fact that this is also the limit of large occupations
 while the reservoir occupation is clamped.
Namely, the terms in the rate equation (\ref{eq:rate_eq}) that describe the intermode rates scale quadratically with the occupation numbers, while the terms describing gain and loss scale only linearly with the occupations. Therefore, the former terms become dominant in the limit of strong pumping.
The homogeneous solution (\ref{eq:final_criterion}) is exactly the solution found in Ref.~\cite{Vorberg2013}
 for a driven-dissipative Bose gas exchanging heat but no particles with its environment.
The only difference is that the total number of particles $N$ occupying the system is not conserved, but determined by gain and loss.

In the limit of very strong pumping, where $\lambda$ becomes large,
 even the mean occupations approach values depending on the rates $R_{ij}$ only.
Namely, asymptotically for strong pumping, we find \cite{Vorberg2013,Vorberg2015}
\begin{eqnarray}
 \nu_i&\simeq& \lambda \nu_i^B  \qquad\qquad\quad\;\, \forall i\in\BES_f,
 \nonumber\\\label{eq:occ_f}
 \nu_i^{(1)}&\simeq& -\frac{\sum_jR_{ij}\nu_j^B}{\sum_jA_{ij}\nu_j^B} \qquad \forall i\notin\BES_f .
\end{eqnarray}

Let us illustrate this finding.
Fig.~\ref{fig:open_vs_close}(a) shows the mode occupations for a toy model where all
 rates are chosen randomly [see figure caption for details].
While the solid lines show the occupations of a model with gain and loss as discussed so far,
 the dashed lines show the ones of a model with fixed particle number (no particle exchange with the environment).
One can see that in the limit of large particle number, which requires large pumping in the open model,
 the same set of states are selected and their occupations are equal in both scenarios.
Here several states become selected even in the limit of large particle number
 since the randomly chosen rates do not obey relation (\ref{eq:db}) and, thus, mimic a situation far from equilibrium \cite{Vorberg2013}.

When the rates $R_{ij}$ describe the contact to a thermal bath of temperature $T$ they obey the relation (\ref{eq:db}).
In this case, the selection criterion (\ref{eq:final_criterion}) predicts ground-state condensation, $\BES_f=\{0\}$.
This is the situation found in exciton-polariton systems, as the one investigated in Ref.~\cite{Galbiati2012}.
Moreover, the mean occupations $n_i=\nu_i^{(1)}$ reached in the limit of strong pumping [Eq.~(\ref{eq:occ_f})]
 approach the equilibrium distribution \cite{Vorberg2013,Vorberg2015}
\begin{align}
 n_i \simeq \frac{1}{e^{\beta(E_i-E_0)}-1}\quad \forall i\ne0
\end{align}
of the Bose-condensed ideal gas, with the chemical potential given by the ground-state energy $E_0$.
Thus, for the exciton-polariton system obeying the condition (\ref{eq:db}),
 equilibrium physics is recovered in the limit of very strong pumping.
This is illustrated in Fig.~\ref{fig:open_vs_close}(b),
 where the only difference to panel (a) is that the rates are chosen to be thermal [Eq.~(\ref{eq:db})]
 by setting
\begin{align}
  \label{eq:thermal_rates}
  R_{ij}^{(b)} =
  \begin{cases}
  R_{ij}^{(a)} & \mbox{for }i\le j\\
  R_{ji}^{(a)}\re^{-\beta (E_i-E_j)} & \mbox{for }i > j,
\end{cases}
\end{align}
with $\beta=1$ and $E_i=i/M$.
The model parameters describing the experiment of Ref.~\cite{Galbiati2012}
 that were used to generate Figs.~\ref{fig:polariton_exp},
 \ref{fig:n_vs_g_and_alpha} an \ref{fig:vs_P} also obey condition (\ref{eq:db}).

\begin{figure*}[ht!]
  \centering
  \includegraphics[width=\linewidth]{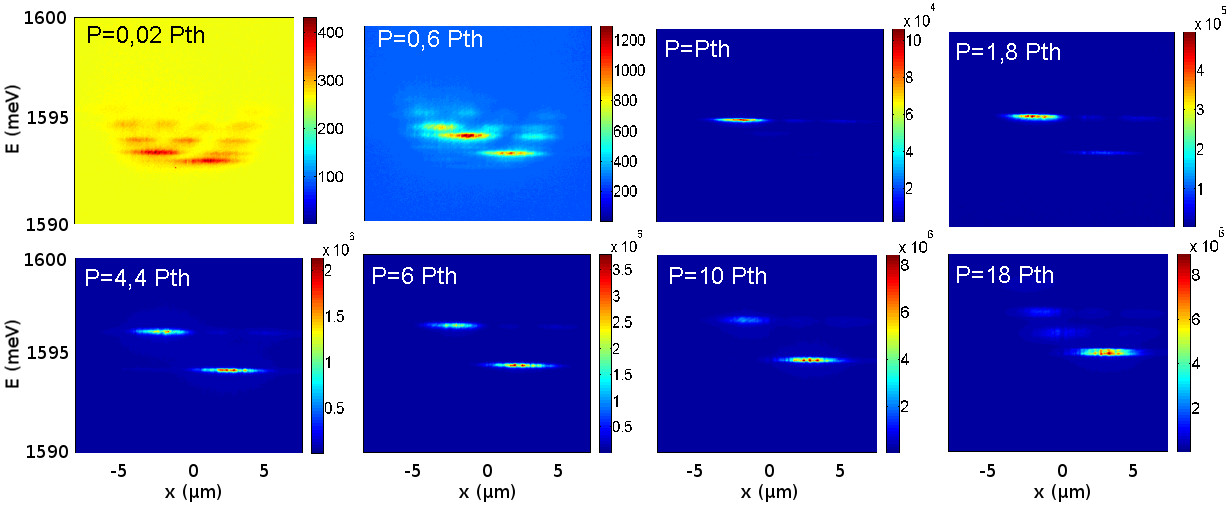}
  \caption{
    Experimental real-space images of the mode occupation in the double-pillar structure investigated in
     Ref.~\cite{Galbiati2010, Galbiati2012}.
    The horizontal axis describes position, the vertical axis energy and the color code intensity.
    One can clearly observe that with increasing pump power a condensate is first
     formed on the left pillar in a mode that corresponds to the third excited state and has a large
     overlap with the reservoir.
    Moreover, for strong pumping ground-state condensation is found.
    However, in agreement with our theory, the two regimes are separated by an intermediate phase where
     both the ground state and third excited state feature a condensate.
    This figure is adapted from Ref.~\cite{Galbiati2010}. Note the different color scale in each panel.}
  \label{fig:exp}
\end{figure*}

\section{Comparison to experiment}
\label{sec:comparison}

Comparing our theory to the experimental results for the double-pillar system by Galbiati
 \emph{et al.}\ \cite{Galbiati2012,Galbiati2010},
 we find good agreement.
In Fig.~\ref{fig:exp} we show experimental data obtained for different pump strengths $P/P_\text{th}$, where $P_\text{th}$ denotes the threshold value above which condensation is observed.
Here the vertical axis denotes energy (obtained from the frequency of the photons extracted from the system) and the horizontal axis position in $x$ direction in real space.
The color code represents the intensity of the emitted light and and is proportional to the particle density (note the different scales in the different panels).
The first state that becomes selected when the pumping is increased is located in the left column,
 where the coupling to the reservoir is strong.
This state corresponds to the third excited state in our example.
This behavior agrees with the prediction that for weak pumping the first condensate is formed
  in the mode with the largest gain-to-loss ratio.
In the limit of strong pumping, eventually the ground state is selected
 despite its location in the right column, where the coupling to the reservoir is weak.
This reflects the prediction that the state reached for strong pumping is independent
 of the details of the gain and loss rates, but depends rather on the energy of the states.
Finally, one can observe that both regimes are separated by a phase occurring at intermediate pump powers,
 where condensates are found in both modes
 \footnote{Note that the very narrow phases [colored green and brown in Fig.~\ref{fig:vs_P}(b)], are not visible in Fig.~\ref{fig:exp}.
 They involve the selection of state 1 and 5 [corresponding to the red and turquoise lines in Figs. \ref{fig:polariton_exp}, \ref{fig:n_vs_g_and_alpha}(b) ans \ref{fig:vs_P}(a) respectively],
  whose occupations remain well below those of the states 0 and 3 when these are selected.
  These narrow phases, i.e., the selection of states 1 and 5, can disappear when choosing slightly different parameters.}.
This behavior, which has not been discussed in Ref.~\cite{Galbiati2012}, is clearly visible in the data presented in Ref.~\cite{Galbiati2010} and reprinted in Fig.~\ref{fig:exp}.
It confirms our theoretical prediction that each transition involves a single state that either becomes or ceases to be selected so that there must be an intermediate phase separating the phase with a single condensate in the third excited mode from the one with a single condensate in the ground state.
The reservoir occupation has not been measured in the experiment.
We would expect that it increases with the pump power before a condensate is formed as well as when two
 condensates are present and that it shows a plateau with respect to the pump power (i.e.\ that it is clamped) whenever a
 single condensate is present.

\section{Conclusion}
\label{sec:conclusion}

We have introduced the concept of Bose selection as a unified description of various forms of nonequilibrum Bose condensation
 in lossy bosonic systems coupled to both a heat bath and a pumped reservoir.
It captures excited-state condensation as an analog of simple lasing,
 equilibrium-like ground-state Bose condensation,
 as well as various situations in between such as fragmented nonequilibrium Bose condensation.
We find that with increasing pump power, the system generically undergoes a sequence of transitions.
Each transition involves a single-particle state that either starts or ceases to be Bose selected.
The first state acquiring a condensate when the pumping exceeds a critical value is
 the one with the largest ratio of pumping to loss.
This intuitive behavior resembles lasing.
Less intuitively, in the limit of very strong pumping,
 Bose condensation is determined by the coupling to the heat bath,
 so that system features a single condensate in the single-particle ground state like in equilibrium.
For intermediate pumping strengths, situations with several condensates have to occur.
Our theory, moreover, predicts that the reservoir occupation forms a plateau
 with respect to the pump power (corresponding to clamping) whenever the system hosts an odd number of Bose condensates.
We compare our theory to experimental data obtained for a system of exciton-polaritons in a double-pillar structure and find good agreement.

The good agreement between theory and experiment suggests that the mechanism of Bose selection is
  robust against weak interactions as they are present in the experiment.
The theoretical investigation of the interacting system constitutes an interesting problem for future research.

The authors thank Jacqueline Bloch, Alexander Leymann, and Alexander Schnell for valuable discussions and Jacqueline Bloch for providing the plots shown in Fig.~\ref{fig:exp}.
D.V.~is grateful for support by the Studienstiftung des Deutschen Volkes.
This work was also supported by the German Research Foundation DFG
 via the Research Unit FOR2414.

\appendix

\begin{widetext}
\section{Relaxation dynamics of a double-pillar polariton system}
\label{app:polartion_rates}

In the following, we describe the parameters that we use to mimic the kinetics of the polariton gas described in Ref.~\cite{Galbiati2012}.
This derivation is motivated by the experiment and follows mostly the numerical simulation discussed in its supplemental material.
The approximative kinetic equations are of the form given by Eq.~\eqref{eq:rate_eq}.
This chapter provides the rates for gain, loss, and the intermode kinetics.
All scattering processes are assisted by either reservoir excitons or phonons. We discuss first the gain rates, then the loss rates, and finally the intermode kinetics.

The processes where two reservoir excitations at energies $E$ and $E'$ (both larger than the bottle-neck energy $E_R$) scatter into the mode $i$ and a reservoir excitation at the higher energy $E'' = E + E' - E_i$ happen at a rate that can be estimated to read,
\begin{align}
  &W_{x} \int \rd x\rd y \int\rd E \rd E' D(E) \rho_R(x,y) n(E) \cdot D(E') \rho_R(x,y) n(E') \cdot |\psi_i|^2 (n_i + 1)\nonumber \\ &\quad\quad\quad\quad\quad\quad\quad D(E'')(1 + \rho_R(x,y)  n(E'')) \nonumber\\
  & \approx W_{x} D_x \NR^2  (\n{i}+1)\int\rd x\rd y  \rho_R^2(x,y) |\psi_i(x,y)|^2 \nonumber\\
  & = \NR^2  (\n{i}+1) g_i^x \quad \mbox{with}\quad g_i^x = W_{x} D_x\int\rd x\rd y  \rho_R^2(x,y) |\psi_i(x,y)|^2.
  \label{eq:app_r_1}
\end{align}
Here is $n(E)=n(0)\re^{-\beta E}$ the exciton distribution, which is Boltzmann-distributed since we assume small occupations in the reservoir only [$n(E) \ll 1$ for all energies $E>E_R$ in the reservoir above the bottleneck energy $E_R$] and $D(E)=D_x$ denotes the exciton's density of states, which is constant since the system is two-dimensional.
We assume that the spatial distribution of the reservoir occupation is given by a Gaussian profile $\rho_R(x, y)$ of width $w$, which is normalized, $\int\rd x\rd y \rho_R(x, y)=1$.
In the second step in Eq.~\eqref{eq:app_r_1}, we used $\NR = \int \rd x\rd y \rd E D(E) \rho_R(x,y) n(E) = \int \rd E  D(E) n(E)$.
Here and henceforth, we neglect, furthermore, the final-state stimulation in the reservoir, $1+ n(E)\approx 1\ \forall E>E_R$.

Scattering can also be assisted by phonons. The process where a reservoir exciton at energy $E$ scatters to the mode at energy
$E_i$ while creating a phonon of energy $E-E_i$ reads
\begin{align}
  &W_{p} \int \rd x\rd y \rd E D(E) \rho_R(x,y) n(E) \cdot |\psi_i|^2 (n_i + 1)\cdot D_p(E - E_i) (1 + n(E - E_i)) \nonumber\\
  & \approx (\n{i}+1) W_{p} \int\rd x\rd y  \rho_R(x,y) |\psi_i(x,y)|^2 = g_i^p (\n{i}+1),
\end{align}
where $D_p(E)=D_p$ is the density of states of the phonons.
The total gain is given by
\begin{align}
  g_i = g_i^x +  g_i^p = W_{x} D_x N_x \int\rd x\rd y  \rho_R^2(x,y) |\psi_i(x,y)|^2 + W_{p} D_p \int\rd x\rd y  \rho_R(x,y) |\psi_i(x,y)|^2
\end{align}
with the parameters $W_xD_x=\SI{2e4}{\per\second}$, $W_pD_p=\SI{1e9}{\per\second}$, $w=\SI{3.0}{\micro\metre}$, and $N_x=100$.

The backward processes from the system to the reservoir
are again either exciton- or phonon-assisted.
The rate of the process where a polariton from mode $i$ and a reservoir excitation at energy $E'' = E+E'-E_i$ scatters to two reservoir excitations at energies $E$ and $E'$ are
\begin{align}
   &W_{x} \int \rd x\rd y \rd E \rd E' |\psi_i|^2 \n{i} \cdot D(E'') \rho_R n(E'') \cdot  D(E) (1+\rho_R n(E)) \cdot D(E')(1+\rho_R  n(E'))\nonumber \\
     &\approx W_{x} \int \rd x\rd y \rd E \rd E' |\psi_i|^2 \n{i} \cdot  D(E'') \rho_R  n(E')\re^{\beta E_i}\re^{-\beta E}\cdot  D(E)  \cdot  D(E')\nonumber \\
   &\approx \NR \frac{D}{\beta}  \n{i}  W_{x} \int\rd x\rd y  \rho_R(x,y) |\psi_i(x,y)|^2 =  \n{i}\ell_i^x.
  %=\n{i} g_i^{xx} \re^{\beta(E_i)} D
\end{align}

The backward process for a phonon assisted scattering from the mode $i$ to the reservoir at energy $E$ by absorption a phonon of energy $E-E_i$ is
\begin{align}
  &W_{p} \int \rd x\rd y \rd E |\psi_i|^2 n_i \cdot D_p(E - E_i) n_p(E - E_i) \cdot  D(E) (1+ \rho_R n(E))   \nonumber\\
  & \approx \n{i} W_{p} D \re^{-\beta E_i } \int\rd x\rd y  \rho_R(x,y) |\psi_i(x,y)|^2  = \n{i} \ell_i^p.
  %= g_i^{xp} \re^{\beta(E_i)} D
\end{align}
The total loss rates are
\begin{align}
  \ell_i = \ell_i^x +  \ell_i^p = N_x  \frac{D}{\beta}    W_{x} \int\rd x\rd y  \rho_R(x,y) |\psi_i(x,y)|^2 + W_{p} D \re^{-\beta E_i } \int\rd x\rd y  \rho_R(x,y) |\psi_i(x,y)|^2.
\end{align}

The intermode kinetics, where a polariton in a state $j$ scatters into the state $i$, is caused either by another polariton or a phonon.
The rate $R_{ij}$ for this process is
\begin{align}
  R_{ij} = \Big[W_{x}N_x\int\rd x\rd y |\langle i|j\rangle|^2\rho_R^2(x,y) + \int\rd x\rd y W_{p}S|\langle i|j\rangle|^2\Big]
    \begin{cases}
      \re^{\beta(E_k-E_i)} &\mbox{for } k < i, \\
      1 & \mathrm{else}.
    \end{cases}
    \label{eq:polartion_R}
\end{align}
Here the first (second) term describes the scattering with an exciton (phonon), respectively.

Furthermore, we assumed the lifetimes to be $\tau_R = \SI{400}{\pico\second}$ and $\tau_i=\SI{20}{\pico\second}$.
\end{widetext}

\section{Implications of the selection criterion}
\label{sec:implications}
In this appendix, we will discuss two implications of the selection criterion
(\ref{eq:selection_criterion}), obtained from requiring positive occupations.
The first one is that the set of selected states $\BES$ contains an even number of selected
states, unless the pump-loss imbalances $W_i$ are fine tuned. The second one is that the
criterion determines a unique set of selected states $\BES$.

Let us first show that without fine tuning of the parameters $W_i$, the number of selected states is even.
Within the subspace of the selected states, from Eq.~(\ref{eq:selection_criterion}) we find
$A^\BES\bnu^\BES +\bW^\BES =0 $, where the superscript indicates projection onto the space of
selected states. Without fine-tuned inhomogeneity $W^\BES_i$, this equation does not possess a
solution, if the matrix $A^\BES$ is singular. And the rate-asymmetry matrix $A^\BES$ is
singular, when $\BES$ contains an odd number of states. Namely, by construction it is skew-symmetric, $A^\BES_{ij}=-A^\BES_{ji}$, implying one eigenvalue zero for odd dimensions.

We will now also proof that the set of selected states obeying the
criterion (\ref{eq:selection_criterion}) is unique, generalizing the approach used in Ref.~\cite{Vorberg2015}.
We will exclude the case of fine-tuned
gain-loss imbalances $W_i$. For that purpose we will show that the assumption that both sets
$\BES_1$ and $\BES_2$ obey the condition (\ref{eq:selection_criterion}) implies that
$\BES_1=\BES_2$. Let $\bnu_{k}$ and $\bmu_{k}$ be the vectors solving
(\ref{eq:selection_criterion}) for $\BES_{k}$. One finds
\begin{align}
\bnu_1^t\bmu_2 =-\bnu_1^t\bmu_2
\label{eq:app_1}
\end{align}
by taking the intermediate steps $\bnu_1^t (A\bnu_2+\bW) = \bnu_1^t A\bnu_2
=\bnu_2^t A^t\bnu_{1} =-\bnu_2^t A\bnu_{1} =-\bnu_2^t(A\bnu_{1}-\bW)$.
Here $\bnu_k^t \bW=0$ has been used, which follows from summing Eq.~(\ref{eq:naive_app}) over $i$.
Since all entries of $\bnu_k$ and $-\bmu_k$ are positive, Eq.~\eqref{eq:app_1} gives
$0\ge \bm{\nu_1^T\mu_2}=-\bm{\nu_1^T\mu_2}\ge 0$ implying
$0=\bm{\nu_1^T\mu_2}=-\bm{\nu_1^T\mu_2}$. Therefore, both $\BES_1\subseteq \BES_2$ and
$\BES_2\subseteq \BES_1$ must hold, so that $\BES_1=\BES_2$.

\bibliography{reference}

\end{document}